\documentclass[aps,prc,reprint,groupedaddress,superscriptaddress]{revtex4-2}  %

\usepackage{amssymb,amsmath,amsfonts}
\usepackage{multirow}
\usepackage{stackengine}
\PassOptionsToPackage{hyphens}{url}
\usepackage{hyperref}
\hypersetup{
    breaklinks=true,
    colorlinks=true,
    urlcolor=blue,
    linkcolor=blue,
    citecolor=blue
}
\makeatletter

\newcommand{\psymbol}[1]{^{\scaleto{\raisebox{1.5pt}{#1}}{5pt}}}
\makeatother
\usepackage{scalerel}
\usepackage{letterspace}
\newcommand{\ROMAN}{\textls*[-100]}
\usepackage[table]{xcolor}

\begin{document}
  \title{$^6$He structure from semiclassical 3D models based on \textit{ab initio} spatial correlations}
  \author{Mengyao Huang}
  \thanks{Contact author: mengyaoh01@gmail.com}
  \thanks{This research was performed while Mengyao Huang was in graduate school at Iowa State University}
  \affiliation{Department of Physics and Astronomy, Iowa State University, Ames, Iowa 50011, USA}
  \author{Tobias Frederico}
  \affiliation{Instituto Tecnol{\'o}gico de Aeron{\'a}utica, DCTA, 12228-900 S{\~a}o Jos{\'e} dos Campos, Brazil}
  \author{Peng~Yin}
  \affiliation{College of Physics and Engineering, Henan University of Science and Technology, Luoyang 471023, China}
  \affiliation{State Key Laboratory of Heavy Ion Science and Technology, Institute of Modern Physics, Chinese Academy of Sciences, Lanzhou 730000, China}
  \affiliation{Department of Physics and Astronomy, Iowa State University, Ames, Iowa 50011, USA}
  \author{James P. Vary}
  \affiliation{Department of Physics and Astronomy, Iowa State University, Ames, Iowa 50011, USA}
  \begin{abstract}
  We construct semiclassical 3D models based on \textit{ab initio} no-core shell model calculated two-body spatial correlations, which represent the preferred spatial configuration of nucleons for the ground state of $^6$He. We demonstrate that the presence of the two valence neutrons distorts and flattens the $\alpha$ core in $^6$He, marking a clear departure from the tetrahedral configuration of free $^4$He, and that the two valence neutrons and the center of mass of the $\alpha$ cluster ($\alpha$-CM) are roughly located at the three distinct vertices of an equilateral triangle. We compute the geometric parameters that commonly characterize $^6$He structure, 1) $\theta_{nn}$, the opening angle of the two valence neutrons with respect to $\alpha$-CM, and 2) $r_{nn}$, the root-mean-square separation between the two valence neutrons. Refining the model with uncertainty estimation, we obtain $\theta_{nn}= 58.5^{+5.4}_{-6.2}$ degrees and $r_{nn}= 4.12^{+0.37}_{-0.43}$ fm, agreeing with those calculated via a phenomenologically modified continuum-discretized coupled-channels method, while an interferometry experiment obtained a marginally higher $r_{nn}$.
  \end{abstract}

\maketitle

\section{Introduction}

The formation of clusters within a nucleus was predicted in the 1930s \cite{doi:10.1098/rspa.1930.0032,RevModPhys.8.82,PhysRev.52.1107}. In particular, $\alpha$-clustering has been proposed for various nuclei, due to the much larger binding energy per nucleon of the $\alpha$-particle (two protons and two neutrons) compared with other light nuclei. For example, the second $0\psymbol{+}$ state 
of $^{12}$C (called the Hoyle state) that plays a key role in stellar nucleosynthesis is often considered a structure composed of three $\alpha$-particles. In addition, the two-$\alpha$ configuration is considered for $^8$Be and the four-$\alpha$ configuration is suggested for $^{16}$O \cite{PhysRevLett.96.192502}. Neutron-rich nuclei that contain a halo structure formed by neutrons outside a closed shell or subshell, such as $^6$He and $^8$He, may also involve $\alpha$-clustering. Experimentally, clustering properties of a nucleus are obtained by combining measurements with theoretical works. For example, Ref.~\cite{MARQUES2000219} relates correlation functions from experimental breakup cross sections with the geometry of the nuclear structure. Others infer structural information by relating the measured or empirically calculated $B(E1)$ with geometrical parameters through the cluster sum rules \cite{PhysRevC.59.1252,SUN2021136072}.

On the theory side, traditional mean-field approaches are challenged to describe nucleons in such a heterogeneous environment that involves clustering. To deal with nuclei with clustering phenomena, one often combines single-nucleon degrees of freedom (DOF) with rotational DOF or vibrational DOF. Methods following this line of thought include the antisymmetrical molecular dynamics (AMD) \cite{kanada-enyo_structure_2007}, the Fermionic molecular dynamics (FMD) \cite{PhysRevLett.98.032501} and the algebraic cluster model \cite{PhysRevLett.113.012502}. Another group of methods extend the DOF of wave functions by adding continuous microscopic-cluster basis states to the discrete basis states of \textit{ab initio} shell model calculations, such as the no-core shell
model/resonating-group method (NCSM/RGM) \cite{quaglioni_no-core_2013}, no-core shell model with continuum (NCSMC) \cite{PhysRevC.97.034332} and the Gamow shell model (GSM) \cite{PhysRevC.84.051304}.

Besides describing a nucleus more accurately with limited computational resources, it is interesting to explore how clustering structures naturally emerge from calculations that do not include cluster information \textit{a priori}. Density profiles extracted from one-body operators with calculation schemes such as the relativistic mean field theory (RMFT) \cite{PhysRevC.71.064308}, the variational Monte Carlo (VMC) \cite{PhysRevC.62.014001}, Monte Carlo shell model (MCSM) \cite{Otsuka:2022bcf}, no-core shell model (NCSM) \cite{cockrell2012ab,PhysRevC.86.034325}, etc. have successfully uncovered important clustering formation patterns in nuclei. The nucleons outside the clusters are often revealed by the difference between the neutron density profile and the proton density profile \cite{cockrell2012ab}, or by subtracting the density profile from that of its isotope (isotone) without excess neutrons (protons) outside of the clusters \cite{Otsuka:2022bcf}. These methods involve assumptions such as the similarity of the neutron and proton density distributions for the clusters, or that adding a few nucleons will not significantly affect the density distribution for each cluster if no new cluster is formed. 

In this work, we explore the feasibility of obtaining a semiclassical 3D model for a nucleus that involves clustering. Although the atomic nucleus is a quantum mechanical object, and the position of each nucleon cannot be fixed in space, we may perform an analysis based on the assumption that among all the possible relative position configurations, there is one dominant in the ground state. Another way to put it is that we can imagine making a sequence of snapshots of the nucleus at a sequence of moments in time, and it will turn out that there is one particular relative position configuration that we have the highest probability of finding. Based on this scenario, we construct 3D models for the relative positions of the nucleons in a nucleus using \textit{ab initio} calculations as input. In practice, we first construct a 3D model of a nucleus with the clustered part inherited from its isotope or isotone containing no unclustered excess nucleons, and then adjust and guide the structure towards a more realistic one using two-body spatial correlations calculated by \textit{ab initio} NCSM. 

This work is based on the work of Ref.~\cite{paper1}. In Ref.~\cite{paper1}, we introduced two-body spatial correlation operators---$r^0$ (called pair-number operator) and $r^2$ (called square-separation operator)---to reveal the spatial correlations for the ground state of $^4$He and $^6$He. We concluded that, in the ground state, the four nucleons of $^4$He are roughly positioned at the four vertices of a tetrahedron. However, complications prevented us from finding the relative positions of the nucleons in $^6$He. In this work, we introduce semiclassical 3D models to analyze these spatial correlations, which allows the exploration of the $^6$He ground state structure in greater detail. 

We first briefly review the two-body spatial correlations relevant to our current discussion (Sec.~\ref{review}). Then we introduce a spin-independent 3D analytical model and summarize its main features (Sec.~\ref{sec_model1}). Later, we improve this model using a 3D numerical model based on spin dependent two-body spatial correlations (Sec.~\ref{sec_model2}). This model is then fine-tuned by expanding the levels of a semiclassical parameter (Sec.~\ref{sec_model3}). Finally, we extract the geometrical parameters and compare them with the results reported in the literature.

\section{Two-body Spatial Correlations}\label{review}
In Ref.~\cite{paper1}, to compute the two-body spatial correlations, we derive the rotationally invariant two-body operator matrix elements projected on the two-body spin $S$ and isospin $z$-components $t_{za}$, $t_{zb}$ in the two-body total angular momentum $J$-coupled basis
\begin{widetext}
\begin{equation}\label{brJM}
\begin{split}
\langle \alpha\beta;J'M'|\mathcal{O}(r)[S,t_{za},t_{zb}]|\gamma\delta;J M\rangle =&\,[\delta_{t_{z\alpha},t_{za}}\delta_{t_{z\beta},t_{zb}}+(1-\delta_{t_{za},t_{zb}})\delta_{t_{z\alpha},t_{zb}}\delta_{t_{z\beta},t_{za}}]\delta_{t_{z\alpha},t_{z\gamma}}\delta_{t_{z\beta},t_{z\delta}}\delta_{J'J}\delta_{M'M}\\
&\times(2S+1)\sqrt{(2j_\alpha + 1)(2j_\beta + 1)(2j_\gamma + 1)(2j_\delta + 1)}\\
&\times\sum_{L }(2L+1)
\begin{Bmatrix}
l_\alpha & l_\beta & L\\
1/2 & 1/2 & S\\
j_\alpha & j_\beta & J
\end{Bmatrix}
\begin{Bmatrix}
l_\gamma & l_\delta & L\\
1/2 & 1/2 & S\\
j_\gamma & j_\delta & J
\end{Bmatrix}\langle n_\alpha l_\alpha n_\beta l_\beta L ||O(r)||n_\gamma l_\gamma n_\delta l_\delta L \rangle,
\end{split}
\end{equation}
\end{widetext}
where $M$ ($M'$) is the $z$ projection of $J$ ($J'$) and $L$ is the two-body total orbital angular momentum. Each Greek symbol represents a single-particle state labeled by a set of quantum numbers, for instance $\alpha\equiv \{n_\alpha, l_\alpha, s_\alpha, j_\alpha, m_{j_\alpha}, t_\alpha, t_{z\alpha}\}$, where the set contains, respectively, the radial quantum number ($n$), orbital angular momentum ($l$), spin ($s=1/2$), total angular momentum and its projection ($j, m_j$), isospin and its projection ($t=1/2, t_z$). In a system of two nucleons, if $O(r) = r^0$, Eq.~(\ref{brJM}) returns the probability of two-nucleon states with $S$, $t_{za}$ and $t_{zb}$; if $O(r) = r^2$, Eq.~(\ref{brJM}) yields the expectation of squared separation between the two nucleons of isospin projections $t_{za}$ and $t_{zb}$, where the two nucleons form two-body spin $S$. For convenience, we note the isospin projections of a pair of nucleons by their corresponding nucleon species, $pp$ (proton-proton), $pn$ (proton-neutron) and $nn$ (neutron-neutron). We introduce a sequence of $\mathcal{S}$ as ($pp0$, $pp1$, $pn0$, $pn1$, $nn0$, $nn1$), where the number in each combination of the sequence is the two-body spin value. We use $\langle r^k\rangle_\mathcal{S}$ to represent the sum of Eq.~(\ref{brJM}) over all possible nucleon pairs in a nucleus, where $O(r) = r^k$.

We employ the \textit{ab initio} NCSM with harmonic-oscillator (HO) basis. We use the Daejeon16 nucleon-nucleon interaction \cite{SHIROKOV201687} obtained from a softened chiral effective field theory interaction, supplemented with the Coulomb interaction between the protons. The HO basis is truncated at $N_{\text{max}}=\sum_i(2n_i+l_i)-N_0$ where $N_0$ is the minimum number of energy quanta required by Pauli exclusion principle. The convergence pattern varies for different $\hbar\Omega$---the energy spacing between shells in the \textit{ab initio} NCSM. We choose $N_{\text{max}}=16$ and $\hbar\Omega=10$ MeV to balance the computational cost with the accuracy needed for the demonstration purpose of our new method (see details in Ref.~\cite{paper1}). The calculated $\langle r^0\rangle_\mathcal{S}$ and $\langle r^2\rangle_\mathcal{S}$ are indicated in Table~II and IV of Ref.~\cite{paper1} and will be used in the following discussions.

\section{3D models of the ground state $^6$He}\label{3d_modeling}
We first set up a 3D analytical model without distorting the $\alpha$ core. I.e., we adopt a tetrahedron structure of the $\alpha$ core (Sec.~\ref{sec_model1}). Then we allow deformations of the $\alpha$ core by constructing a spin-dependent 3D numerical model (Sec.~\ref{sec_model2}). We also allow for deviations caused by ambiguities in determining the rms separation for each pair of the nucleons in these semiclassical models (Sec.~\ref{sec_model3}).

\subsection{3D analytical model}\label{sec_model1}

\begin{figure}[htp]
\centering
\includegraphics[width=\linewidth]{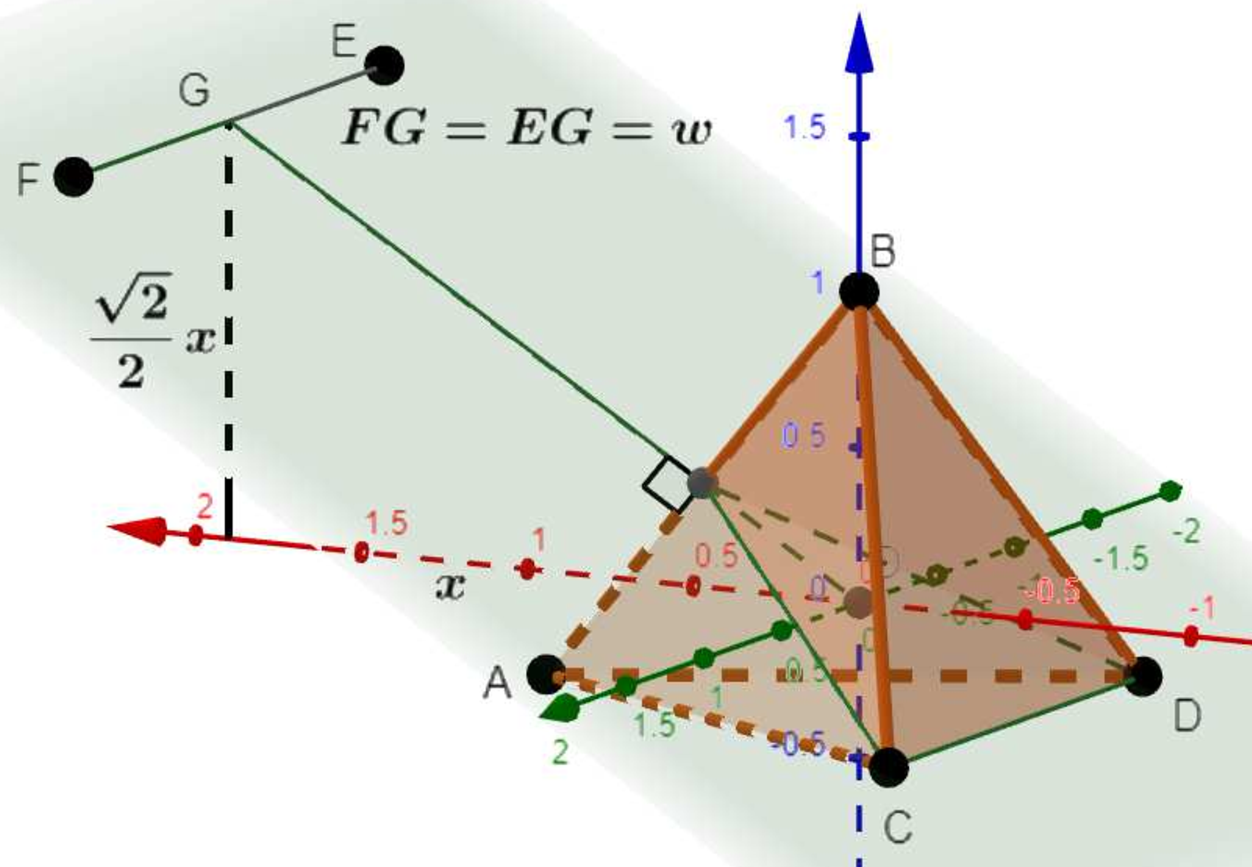}
\caption{The analytical model of $^6$He parametrized by $x$ and $w$. The origin is at the center of the tetrahedron. $EF$ is parallel to $CD$ and perpendicular to $AB$. The numerical scales on the axes are in units of $r_0$. The green tilted plane is defined by the origin and the locations of the two valence neutrons.}
\label{fig:he6_model_design}
\end{figure}

We construct a 3D analytical model of $^6$He by adding two neutrons to a tetrahedron $\alpha$ core. Since the ground state of $^4$He has a near-tetrahedron structure \cite{paper1}, it is natural to retain this structure for the $\alpha$ core in the ground state of $^6$He, as a starting point for a more complicated structure. The geometrical structure of this model is illustrated in Fig.~\ref{fig:he6_model_design}. The four vertices of the tetrahedron represent the locations of the four $\alpha$ core nucleons in $^6$He, which are labeled as $A$, $B$, $C$ and $D$. The center of the tetrahedron is placed at the origin. The distance from any of the tetrahedron vertices to the center of tetrahedron is $r_0$. The locations of the two valence neutrons are labeled as $E$ and $F$, respectively. $|EF|=2w$ and $EF$ parallels to the $y$ axis with $x$ being the $x$-coordinate of $E$ or $F$. The perpendicular bisector of $EF$ connecting to the origin also perpendicular bisects $AB$. This arrangement restricts the locations of the valence neutrons to the plane defined by $C$, $D$ and the origin, illustrated in Fig.~\ref{fig:he6_model_design}. The resulting coordinates of the six nucleons are $A(\frac{2\sqrt{2}}{3}r_0,0,-\frac{1}{3}r_0)$, $B(0,0,r_0)$, $C(-\frac{\sqrt{2}}{3}r_0,\frac{\sqrt{6}}{3}r_0,-\frac{1}{3}r_0)$, $D(-\frac{\sqrt{2}}{3}r_0,-\frac{\sqrt{6}}{3}r_0,-\frac{1}{3}r_0)$, $E(x,-w,\frac{\sqrt{2}}{2}x)$ and $F(x,w,\frac{\sqrt{2}}{2}x)$. Although in this simplified model, we try to preserve the geometric symmetry in $^6$He to the same extent as in $^4$He, the two neutrons outside the tetrahedron are not equidistant from all four $\alpha$ core nucleons. I.e., in Fig.~\ref{fig:he6_model_design}, $|EA|=|EB|=|FA|=|FB|$, but $|ED|=|FC|\neq |EC|=|FD|$. The squared separations between nucleons are thus parameterized as
\begin{equation}\label{eq:analytical_parameterized}\begin{split}
|EA|^2&=|EB|^2=|FA|^2=|FB|^2\\
&=(x-\frac{2\sqrt{2}}{3}r_0)^2+w^2+(\frac{\sqrt{2}}{2}x+\frac{r_0}{3})^2,\\
|EC|^2&=|FD|^2\\
&=(x+\frac{\sqrt{2}}{3}r_0)^2+(w+\frac{\sqrt{6}}{3}r_0)^2+(\frac{\sqrt{2}}{2}x+\frac{r_0}{3})^2,\\
|ED|^2&=|FC|^2\\
&=(x+\frac{\sqrt{2}}{3}r_0)^2+(w-\frac{\sqrt{6}}{3}r_0)^2+(\frac{\sqrt{2}}{2}x+\frac{r_0}{3})^2,\\
|EF|^2&=4w^2.
\end{split}\end{equation}

To connect this model with the calculated \textit{ab initio} two-body spatial correlations, we introduce 
\begin{equation}\label{eq:analytical_rnp_b}\begin{split}
a&\equiv\langle r^2\rangle_{np}^{(^6\text{He})}-\langle r^2\rangle_{np}^{(^4\text{He})},\\
b&\equiv\langle r^2\rangle_{nn}^{(^6\text{He})}-\langle r^2\rangle_{nn}^{(^4\text{He})},
\end{split}\end{equation}
where $a=73.4800$~fm$^2$ and $b=95.8616$~fm$^2$ according to Table IV of Ref.~\cite{paper1} regardless of $S$. Two values of $r_0$ are tested. One is $r_0=1.51$~fm, the calculated $^4$He point-nucleon radius \cite{paper1}. The other is $r_0=1.89$~fm corresponding to a swelled $\alpha$ core indicated by the expanded point-proton radius for $^6$He \cite{paper1}. A total of three cases are generated for this ``nearly symmetrical'' analytical model. In each case, $a$ ($b$) of Eq.~(\ref{eq:analytical_rnp_b}) is expressed as the sum of squared separations for $np$ ($nn$) pairs  partially or entirely outside the $\alpha$ core. For convenience, in the following text, the unparenthesized (parenthesized) value corresponds to the results for $r_0=1.51$ (1.89)~fm. %

\begin{figure*}[htp]
  \centering
  \begin{tabular}[c]{c @{\hspace{0ex}} c}
    \includegraphics[width=0.45\textwidth]{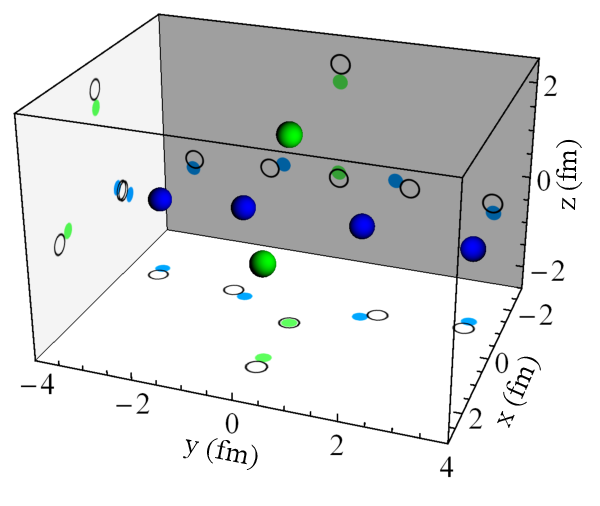}
    &
    \includegraphics[width=0.45\textwidth]{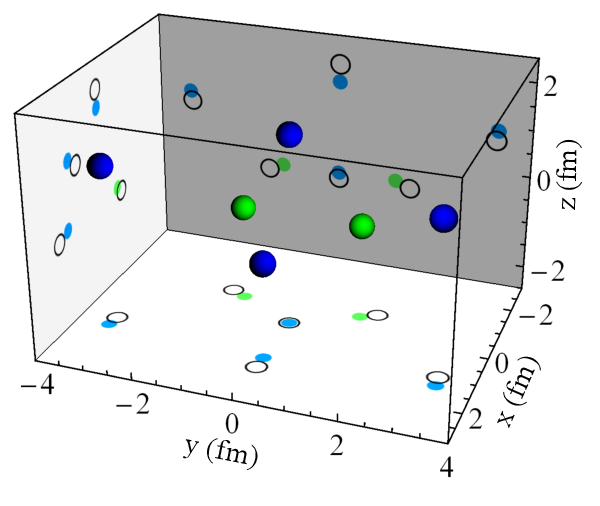}
    \\[-2mm]
    (a) & (b)
    \\[2mm]
    \includegraphics[width=0.45\textwidth]{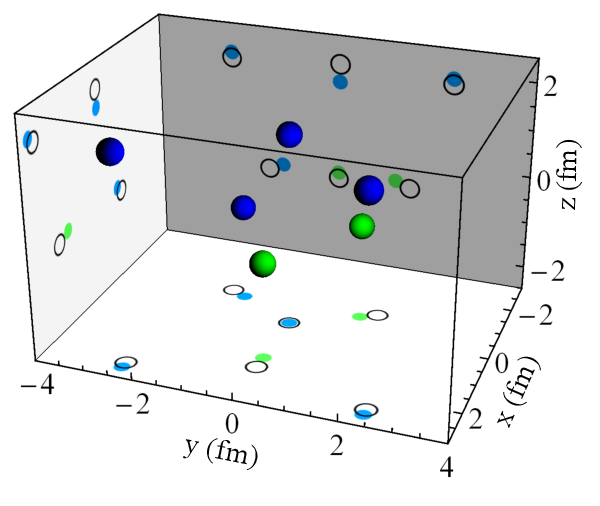}
    &
    \includegraphics[width=0.45\textwidth]{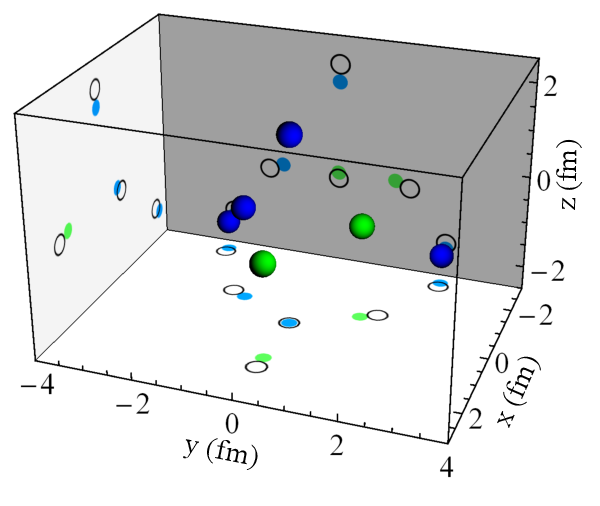}
    \\[-2mm]
    (c) & (d)
  \end{tabular}
  \caption{$^6$He 3D analytical model for Case (a) \ROMAN{I}, (b) \ROMAN{II}, (c) \ROMAN{III}-first solution, and (d) \ROMAN{III}-second solution. The nucleon positions (color balls) and their projections (color dots) on the three perpendicular walls of a rectangle box are shown. For $r_0=1.51$~fm, the green (blue) color represents protons (neutrons). For $r_0=1.89$~fm, only the projections are shown, indicating by a circle nearby the color dot for each nucleon.}
  \label{fig:cases}
\end{figure*}

\begin{itemize}
\item Case~\ROMAN{I}: $A$ and $B$ are protons%
\begin{enumerate}
\item[i)]
$4|EA|^2=a$;
\item[ii)]
$2|EC|^2+2|ED|^2+|EF|^2=b$.
\end{enumerate}
Substituting Eq.~(\ref{eq:analytical_parameterized}) into the above equations, we find
\begin{equation}\begin{split}\label{case1}
x&=\sqrt{2}r_0\pm\frac{2\sqrt{3}}{3}\sqrt{r_0^2-\frac{b}{8}+\frac{a}{4}},\\
w^2&=\frac{b}{4}-\frac{a}{4}-2\sqrt{2}xr_0,
\end{split}\end{equation}
which leads to two solutions of ($x$, $w$). $w^2>0$ excludes the solution corresponding to the positive sign in the second term of $x$ of Eq.~(\ref{case1}). The surviving solution is $x=-1.264$ ($-0.971$)~fm and $w=3.316$ (3.284)~fm. The resulting nucleon positions are illustrated in Fig.~\ref{fig:cases}(a). %
The valence neutrons are very close to the $\alpha$ core neutrons with a minimum separation of 2.19 (1.74)~fm, which is less than the average separation of nucleons within the $\alpha$ core, 2.47 (3.09)~fm, so each of the two valence neutrons strongly attracts one of the $\alpha$-core neutrons such that the $\alpha$ core no longer exists. This contradicts the reality and therefore Case~\ROMAN{I} is not valid.

\item Case~\ROMAN{II}: $C$ and $D$ are protons%
\begin{enumerate}
\item[i)]
$2|EC|^2+2|ED|^2=a$;
\item[ii)] 
$4|EA|^2+|EF|^2=b$.
\end{enumerate}
It follows that
\begin{equation}\begin{split}\label{case2}
x&=-\sqrt{2}r_0\pm\frac{2\sqrt{3}}{3}\sqrt{r_0^2-\frac{b}{8}+\frac{a}{4}},\\
w^2&=\frac{b}{4}-\frac{a}{4}+2\sqrt{2}xr_0,
\end{split}\end{equation}
where the solution corresponding to a negative second term of $x$ is discarded due to $w^2>0$. We obtain $x=1.264$ (0.971)~fm and $w=3.316$ (3.284)~fm. The resulting positions of nucleons are visualized in Fig.~\ref{fig:cases}(b). Each valence neutron is %
3.60 (3.63)~fm from each of the $\alpha$ core neutrons, much larger than in Case~\ROMAN{I}. %
The smallest separation between the valence neutrons and the $\alpha$ core nucleons is 3.19 (2.87)~fm. For $r_0=1.51$~fm, the separation between any $\alpha$ core nucleon and any valence neutron is larger than the separation of nucleons within the $\alpha$ core. The same applies to $r_0=1.89$~fm except for the two-nucleon separation of one $np$ pair, 2.87~fm, that is moderately smaller than the $\alpha$ core two-nucleon separation 3.09~fm. The two valence neutrons form an angle of 118.7 (128.5)$^\circ$ with respect to the center of the $\alpha$ core. This obtuse angle implies a ``cigar'' configuration. However, as suggested by the above exception, this configuration does not strongly support $\alpha$ clustering.

\item Case~\ROMAN{III}: $A$ and $C$ are protons%
\begin{enumerate}
\item[i)] 
$2|EA|^2+|EC|^2+|ED|^2=a$;
\item[ii)] 
$2|EA|^2+|EC|^2+|ED|^2+|EF|^2=b$.
\end{enumerate}
We get
\begin{equation}\begin{split}\label{case3}
x&=\pm \frac{\sqrt{6}}{3}\sqrt{\frac{a}{2}-r_0^2-\frac{b}{4}},\\
w^2&=\frac{b}{4}-\frac{a}{4},
\end{split}\end{equation}

where $w=2.365$~fm and $x=\pm 2.645$ ($\pm 2.477$)~fm. For the positive-$x$ solution, the positions of the nucleons are illustrated in Fig.~\ref{fig:cases}(c). Each valence neutron is 4.01 (3.85)~fm from the center of the $\alpha$ core. The smallest separation between the valence neutrons and the $\alpha$ core nucleons is 3.57 (3.43)~fm, larger than in Case~\ROMAN{II}. The valence neutrons form an angle of 68.8 (71.6)$^{\circ}$ with respect to the center of the $\alpha$ core, which is less than $90^{\circ}$ for two noncorrelated neutrons. This provides a dineutron picture of the halo structure. %

For the negative-$x$ solution, the positions of the nucleons are shown in Fig.~\ref{fig:cases}(d). Each valence neutron is again 4.01 (3.85)~fm from the center of the $\alpha$ core. However, the smallest separation between the valence neutrons and the $\alpha$ core nucleons is only 2.62 (2.11)~fm, comparable to (less than) the two-nucleon separation inside the $\alpha$ core, so we discard this solution for a similar reason as in Case~\ROMAN{I}.

\end{itemize}

We conclude from the above analysis that it is feasible to form a semiclassical 3D structure for the ground state of $^6$He by matching our spin-independent \textit{ab initio} calculated results with the nearly symmetrical analytical model. We obtain a favorable halo picture, such as the one illustrated in Fig.~\ref{fig:cases}(c). %
This gives us the confidence to pursue a more refined halo picture of $^6$He, where we no longer keep the $\alpha$ core in a tetrahedron shape. Instead, we use Case~\ROMAN{III}'s first solution as an initial trial solution for the relative position of nucleons, and let it evolve in numerical iteration process to find the solution of the numerical equations of squared separation in the next subsection (Sec.~\ref{sec_model2}). %

\subsection{3D numerical model for the dominant two-body spin configuration}\label{sec_model2}

\begin{figure*}[t]
  \centering
  \begin{tabular}[c]{cc}
    \includegraphics[width=0.4\textwidth]{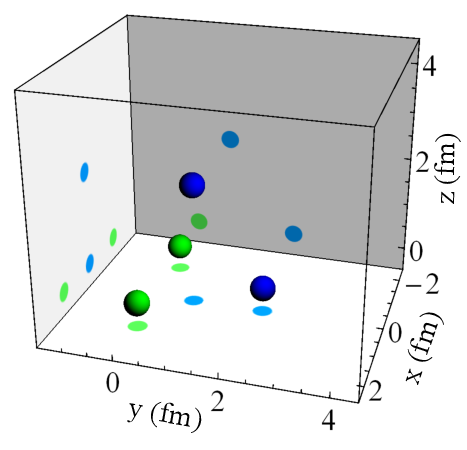}
    &
    \includegraphics[width=0.4\textwidth]{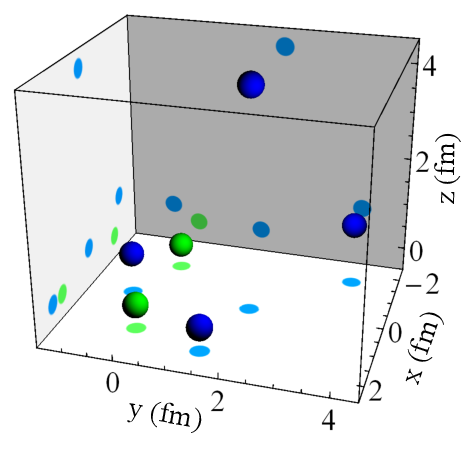}
    \\[-2mm]
    (a) & (b)
  \end{tabular}
  \caption{The 3D numerical model for (a) $^4$He and (b) $^6$He in Sec.~\ref{sec_model2}. The nucleon positions (color balls) and their projections (color dots) on the three perpendicular walls of a rectangle box are shown. The green color represents the protons.}
  \label{fig:model_spin}
\end{figure*}

The squared separation between the $i$-th and $j$-th nucleons is
\begin{equation}
\label{rij}
\langle r^2 \rangle_{ij}=(x_i-x_j)^2+(y_i-y_j)^2+(z_i-z_j)^2.
\end{equation}
To find $\langle r^2 \rangle_{ij}$ for the dominant two-body spin configuration, we express $\langle r^0 \rangle_\mathcal{S}$ as
\begin{equation}\label{r0_rewrite}
\langle r^0 \rangle_\mathcal{S}=\sum_k P_kB_{k\mathcal{S}},
\end{equation}
where $k$ is the index of possible two-body spin configurations. $B_{k\mathcal{S}}$ is the number of nucleon pairs in the $\mathcal{S}$ (defined by isospin projections and spin in Sec.~\ref{review}) channel and the $k$-th two-body spin configuration. $P_k$ represents the probability of the $k$-th two-body spin configuration, and $P_0$ is the highest one (see Appendix~\ref{appendix:configurations} for further explanations). Since the mean square separation between two nucleons with given $\mathcal{S}$ in a system of nucleons is defined as
$\overline{\langle r^2 \rangle}_\mathcal{S} \equiv
 \langle r^2\rangle_\mathcal{S}/\langle r^0\rangle_\mathcal{S}$ \cite{paper1}, we rewrite Eq.~\ref{r0_rewrite} as
\begin{equation}\label{r2_rewrite}
\overline{\langle r^2 \rangle}_\mathcal{S}\cdot\sum_{k}P_kB_{k\mathcal{S}}=\langle r^2 \rangle_\mathcal{S}.
\end{equation}
The squared separation for the dominant two-body spin configuration with a given $\mathcal{S}$ is
\begin{equation}\begin{split}\label{rijfromr2}
\overline{\langle r^2 \rangle}_\mathcal{S}\cdot B_{0\mathcal{S}}=&\sum_{i'<j'}^{N_\mathcal{S}}\langle r^2 \rangle_{i'j'}[\mathcal{S}]\\=&\sum_\epsilon\sum_{i''<j''}^{N_{\mathcal{S},\epsilon}}\langle r^2 \rangle_{i''j''}[\mathcal{S},\epsilon]\\
=&\sum_\epsilon\overline{\langle r^2 \rangle}_{\mathcal{S},\epsilon}N_{\mathcal{S},\epsilon},
\end{split}\end{equation}
where we introduce a semiclassical parameter $\epsilon$. Here, $\epsilon$ encodes the clustering information. More specifically, $\epsilon=1$ is for nucleon pairs with both nucleons within the $\alpha$ core, otherwise $\epsilon=2$. From Ref.~\cite{paper1}, $\overline{\langle r^2\rangle}$ = (6.85424, 11.1484, 12.3921, 12.1607, 13.5121, 20.1827)~fm$^2$ and $B_{0}$ = (1, 0, 2, 6, 3, 3), the components of which are ordered by our defined sequence of $\mathcal{S}$ given in Sec.~\ref{review}. If pair $ij$ belongs to the channel defined by $\mathcal{S}$ and $\epsilon$,
\begin{equation}\label{rij_get}
\langle r^2\rangle_{ij}\approx\overline{\langle r^2\rangle}_{\mathcal{S},\epsilon}.
\end{equation}
We introduce three parameters $q_1$, $q_2$ and $q_3$ to represent the unknown $\overline{\langle r^2\rangle}_{\mathcal{S},1}$ values in Eq.~(\ref{rijfromr2}), see Table~\ref{table_numer_m}. The number of equations of Eq.~(\ref{rij}), where $1\leq i<j\leq A$ for an $A$-body system, equals the total number of pairs, $A(A-1)/2$. We solve this set of equations numerically using SciPy.optimize.fsolve with initial relative locations provided by the valid solution of the analytical model Case~\ROMAN{III} in Sec.~\ref{sec_model1}. We obtain $q_1=11.242$~fm$^2$, $q_2=2.1699$~fm$^2$ and $q_3=7.0830$~fm$^2$. The resulting relative locations of the $^6$He nucleons for the dominant two-body spin configuration are illustrated in Fig.~\ref{fig:model_spin}(b). 

\begin{table}[htp] \centering
\renewcommand{\arraystretch}{1.2}
\setlength{\extrarowheight}{1pt}
\setlength{\tabcolsep}{4.2pt}
\caption{The two-nucleon squared separation {[fm$^2$]} for $^6$He dominant two-body spin configuration, categorized by $\mathcal{S}$ and $\epsilon$. The bracketed expressions indicate the changes after the modification of the numerical model.}
\label{table_numer_m}
\begin{tabular}{ccccc} 
\hline\hline
$\mathcal{S}$
& \addstackgap[2pt]{$\overline{\langle r^2 \rangle}_\mathcal{S}\cdot B_{0\mathcal{S}}$}
&$\epsilon=1$ 
&$\epsilon=2$
&$\epsilon=3$\\
\hline
$pp0$ & 6.8542 & 6.8542 & - & -\\
$pp1$ & 0.0000 & - & - & -\\
$pn0$ & 24.784 & $q_1$ & $24.784-q_1$ & - \\
$pn1$ & 72.964 & $3q_2$ & $72.964-3q_2$ & - \\
\multirow{2}{*}{$nn0$} & \multirow{2}{*}{40.536} & \multirow{2}{*}{$q_3$} & $40.536-q_3$ & - \\
&&&($\frac{40.536-q_3}{2}-\xi$)&$(\frac{40.536-q_3}{2}+\xi$)\\
$nn1$ & 60.548 & - & 60.548 & - \\
\hline\hline
\end{tabular}
\end{table}

To our surprise, although starting with a tetrahedron structure, the $^6$He $\alpha$ core is almost squeezed to a plane when the numerical solution of Eq.~\ref{rij} is found. To check the validity of the numerical method, we build a similar numerical model for $^4$He. The resulting nucleon locations are shown in Fig.~\ref{fig:model_spin}(a). As expected, without the influence of the two additional neutrons, the bare $\alpha$ particle essentially preserves the tetrahedron structure after this numerical method is performed.

\begin{figure}[h]
  \centering
\includegraphics[width=0.35\textwidth]{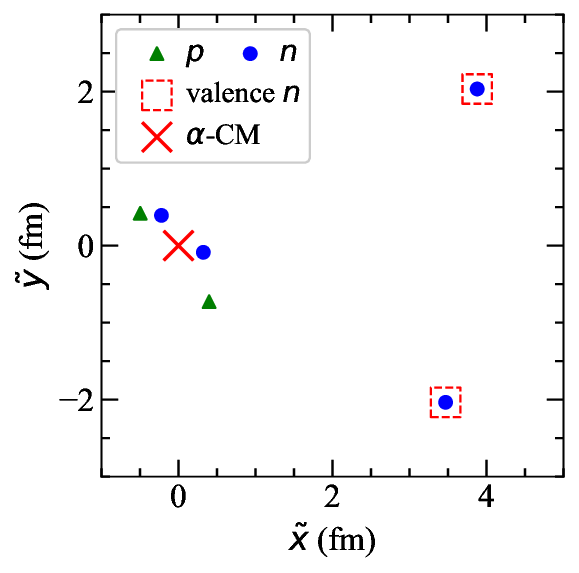}
  \caption{The nucleon positions projected on the plane defined by $\alpha$-CM and the locations of the two valence neutrons, where $\tilde{x}$ and $\tilde{y}$ are the coordinates of the plane with $\alpha$-CM at the origin and $\tilde{x}$ points to the midpoint of the positions of the two valence neutrons.}
  \label{fig:alpha-2n}
\end{figure}

To obtain the geometrical parameters quantifying the correlation of the two valence neutrons \cite{PhysRevLett.125.252501}, we project the nucleon positions onto a plane defined by the center of mass of the $\alpha$ core ($\alpha$-CM) and the two valence neutrons, shown in Fig.~\ref{fig:alpha-2n}. The opening angle between the valence neutrons with respect to $\alpha$-CM is $\theta_{nn}=58.1^{\circ}\lesssim60^{\circ}$. The separation between the valence neutrons $r_{nn}=4.09$~fm. The separations between $\alpha$-CM and the two valence neutrons are 4.02~fm and 4.38~fm, respectively. Therefore, $\alpha$-CM and the two valence neutrons form a nearly equilateral configuration.

It is interesting to note that if we ignore the difference in the shape of the $\alpha$ core, the numerical model is, in some sense, a mixture of the analytical model Case~\ROMAN{II} and \ROMAN{III}. In Case~\ROMAN{II}, $\alpha$-core $nn$ forms a 90$^\circ$ angle with the valence $nn$. In Case~\ROMAN{III}, $\alpha$-core $pp$ and $nn$ form the same 60$^\circ$ angles with the valence $nn$. In this numerical model, the angle between the $\alpha$-core $pp$ and the valence $nn$ is  66.3$^\circ$ (similar to Case~\ROMAN{III}) and the angle between the $\alpha$-core $nn$ and the valence $nn$ is  80.9$^\circ$ (close to Case~\ROMAN{II}).

\subsection{Modification of the 3D numerical model and uncertainty estimation}\label{sec_model3}

We modify the numerical model by adding a third level to the semiclassical parameter, $\epsilon=3$. This is because in the original numerical model, the value in $nn0$ with $\epsilon=2$ is shared by two $nn$ pairs with very different characteristics considering their semiclassical properties. One pair is formed by an $\alpha$-core neutron and a valence neutron. The correlation between the two nucleons of this pair is a correlation between shells. The other pair is formed by the two valence neutrons in the same shell. We introduce a parameter $\xi$ to account for this difference, shown in Table~\ref{table_numer_m}. As in Sec.~\ref{sec_model2}, we solve the set of equations of Eq.~(\ref{rij}) for a sequence of values of $\xi$ with an increment of 0.5~fm$^2$ from $\xi=-5$ to 5 fm$^2$. From Eq.~(\ref{rij_get}), the rms separation $r_{ij}\approx\sqrt{\overline{\langle r^2 \rangle}_{\mathcal{S},\epsilon}}\equiv r_{\mathcal{S},\epsilon}$, if pair $ij$ belongs to the ($\mathcal{S},\epsilon$) channel. %
We consider a reasonable uncertainty range of $\xi$ based on the following hypothetical criteria:
\begin{enumerate} 
\item The rms separation between any two nucleons of the $\alpha$ core should not be larger than the rms separation between any $\alpha$-core nucleon and any valence neutron.%
\item For pairs with identical isospin projections ($nn$ or $pp$), the rms separation for $S=0$ should be smaller than $S=1$ in the ground state. This applies to both ground state $^4$He and $^6$He, as seen in Table~IV of Ref.~\cite{paper1}.
\end{enumerate}

\begin{figure}[htp!]
\centering
\includegraphics[width=\linewidth]{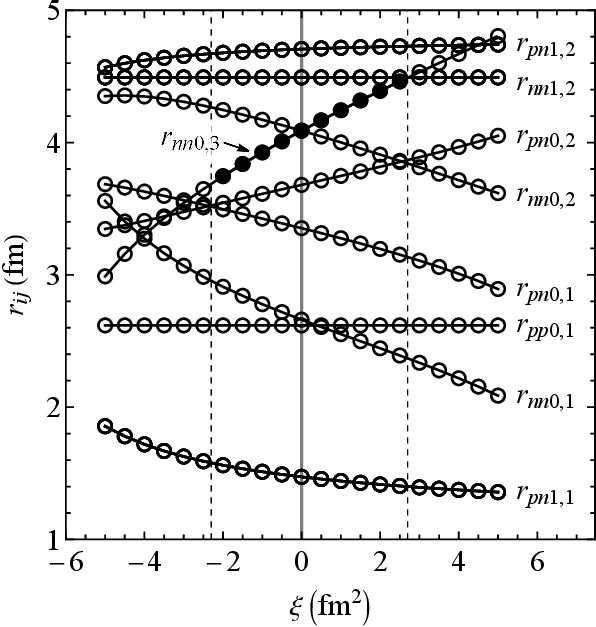}
\caption{
The two-nucleon separation $r_{ij}$ ($i<j$ runs through all nucleon pairs) in the modified 3D numerical model for $^6$He, categorized by $r_{\mathcal{S},\epsilon}$. See text for explanations. The filled circles indicate the values used to estimate $r_{nn}$.}
\label{fig:he6_separation}
\end{figure}

\begin{figure}[htp!]
\centering
\includegraphics[width=\linewidth]{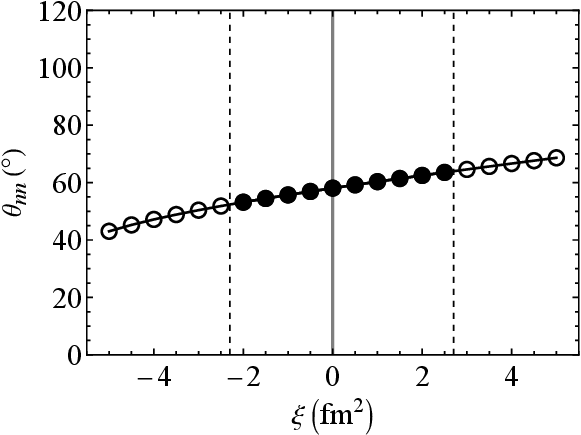}
\caption{
The opening angle $\theta_{nn}$ for $^6$He. The dashed lines delineate the bounds of the reasonable range of $\xi$. The filled circles indicate the values used to estimate $\theta_{nn}$.}
\label{fig:he6_numer_a}
\end{figure}

The resulting $r_{ij}$ categorized by $r_{\mathcal{S},\epsilon}$ are shown in Fig.~\ref{fig:he6_separation}. The solid vertical line at $\xi=0$ corresponds to the original 3D numerical model in Sec.~\ref{sec_model2}. 
When $\xi<-2.3\text{~fm}^2$, $r_{np0,2}$ is less than $r_{np0,1}$, the first criterion is not satisfied. The second criterion is not met when $r_{nn0,3}$ exceeds $r_{nn1,2}$ at $\xi>2.7\text{~fm}^2$. Therefore, we estimate the reasonable range of $\xi$ as $-2.3\text{~fm}^2\leq\xi\leq2.7\text{~fm}^2$, indicated by the region within the dashed lines in Fig.~\ref{fig:he6_separation}. We obtain $r_{nn}=r_{nn0,3}=4.12_{-0.43}^{+0.37}$~fm, where we estimate the central value as the mean value within the reasonable range of $\xi$, and the upper (lower) uncertainty as the deviation from the highest (lowest) value to the mean within the range. We extract $\theta_{nn}$ using the same rule and get $\theta_{nn}=58.5_{-6.2}^{+5.4}$~degrees, shown in Fig.~\ref{fig:he6_numer_a}. 

We compare our $\theta_{nn}$ and $r_{nn}$ values with those reported in the literature in Table~\ref{compare}. Both values obtained by the empirically modified four-body continuum-discretized coupled-channels (CDCC) \cite{SUN2021136072} method are consistent with our results. This method utilizes the relationship between the total $B(E1)$ and the geometrical parameters, where the $B(E1)$ spectrum is extracted from the CDCC fit to the experimental cross sections with tunable parameters, assuming that all events are elastic. However, $\theta_{nn}$ deduced from the original microscopic CDCC \cite{SUN2021136072} without tunable parameters is about $10^\circ$ greater than both modified CDCC and our results. %
We further compare our $r_{nn}$ result with that obtained from a two-neutron interferometry experiment \cite{MARQUES2000219}. The maximum value of $r_{nn}$ within the uncertainty range of both our work and the modified CDCC are approximately 0.2 fm smaller than the minimum value obtained via the interferometry technique, which is believed to be more sensitive to continuum states in the dissociation than to the ground state \cite{orr2008neutron}. 

\begin{table}[htp!]
\centering
\renewcommand{\arraystretch}{1.2}
\caption{$^6$He geometrical parameters}
\label{compare}
\begin{tabular}{ccc} 
 \hline  \hline
 & $r_{nn}$ (fm)
 & $\theta_{nn}$ ($^{\circ}$) 
 \\ \hline 
CDCC/modified CDCC \cite{SUN2021136072}
& N.A./4.1(7) & 68/$56^{+9}_{-10}$
\\Interferometry \cite{MARQUES2000219} & $5.9{\pm}1.2$ & N.A. \\
This work & $4.12_{-0.43}^{+0.37}$ & $58.5_{-6.2}^{+5.4}$
 \\\hline \hline
\end{tabular}
\end{table}

In addition, we check that the 3D numerical model reproduces the point-proton radius $r_p$ and point-neutron radius $r_n$ in the original \textit{ab initio} calculations, see Appendix~\ref{appendix:r_cluster}. The resulting locations of the nucleons are not far away from those obtained in Sec.~\ref{sec_model2}, illustrated in Appendix~\ref{appendix:3d_model}.

\section{Conclusion and Discussion}\label{conclusion}

We use a semiclassical 3D modeling method to reveal the structure of the dominant two-body spin configuration for the ground state of $^6$He, based on two-body spatial correlations $r^0$ and $r^2$ calculated by the \textit{ab initio} NCSM. We obtain a nearly planar shape for the $\alpha$ cluster in $^6$He. This unusual shape contradicts our intuition that the $\alpha$ cluster should resemble a $^4$He nucleus. This is due to the fact that the correlated halo neutrons significantly distort the $\alpha$-nucleon system from its free-space shape. I.e., we obtain a classical version of a significant ``core-polarization'' effect. In addition, we found a nearly equilateral triangle configuration for the center of mass of the $\alpha$ cluster with the two valence neutrons.

It is convenient to deduce geometrical parameters using this semiclassical 3D modeling method. The opening angle of the two valence neutrons with respect to the center of mass of the $\alpha$ cluster $\theta_{nn}$ is $58.5_{-6.2}^{+5.4}$~degrees and the rms separation between the two valence neutrons $r_{nn}$ is $4.12_{-0.43}^{+0.37}$~fm, both of which are consistent with the results obtained by a phenomenologically modified CDCC method \cite{SUN2021136072}. By comparing our $r_{nn}$ with the one deduced from an interferometry experiment \cite{MARQUES2000219}, we find that the upper limit of our $r_{nn}$ is 0.2~fm lower than the minimum value for the uncertainty range obtained in this experiment. This deviation likely stems from the interferometer's sensitivity to the continuum states. %

\vspace{2mm}
\begin{acknowledgements} 
This work was supported in part by the US Department of Energy (DOE) under Grant No.~DE-FG02-87ER40371, No.~DE-SC0018223 (SciDAC-4/NUCLEI), No.~DE-SC0023495 (SciDAC-5/NUCLEI), No.~DE-SC0023707 (Quantum Horizons---NuHaQ). T.F. thanks the hospitality of the Department of Physics and Astronomy of Iowa State University and Fulbright Brazil for the financial support during his visits in 2018 and 2019. T.F. also thanks the National Council for Scientific and Technological Development (CNPq) Grant No.~306834/2022-7, the S\~ao Paulo Research Foundation (FAPESP) Grant No.~2023/13749-1 and No. 2024/17816-8, and the INCT-FNA Project No.~408419/2024-5. P.Y. is supported by National Natural Science Foundation of China Grant No. 12675152, Natural Science Foundation (Henan, China)
under Grant No.~252300421486, and by the State Key Laboratory of Heavy Ion Science and Technology, Institute of Modern Physics, Chinese Academy of Sciences Grant No. HST2026CO09. We thank the Institute for Nuclear Theory at the University of Washington for its kind hospitality and stimulating research environment. This research was supported in part by the INT's U.S. Department of Energy Grant No.~DE-FG02-00ER41132.
\end{acknowledgements}

\appendix

\section{Two-body spin configurations}\label{appendix:configurations}

In $^4$He, the calculated $\langle r^0\rangle$ = (0.976772, 0.023228, 0.979111, 3.020890, 0.976774, 0.023226) \cite{paper1}. $B_0$ = (1, 0, 1, 3, 1, 0) obtained by rounding $\langle r^0\rangle_\mathcal{S}$ to the nearest integers. Since the round-off errors are small relative to one, $B_0$ represents the dominant two-body spin configuration and $P_0$ is the highest probability. Other possible two-body spin configurations are $B_1$ = (1, 0, 1, 3, 0, 1), $B_2$ = (0, 1, 1, 3, 0, 1), $B_3$ = (1, 0, 0, 4, 1, 0), etc.. Similarly, the dominant two-body spin configuration for $^6$He is $B_0$ = (1, 0, 2, 6, 3, 3) based on $\langle r^0\rangle$ = (0.967578, 0.032422, 1.977430, 6.022570, 2.865420, 3.134580) \cite{paper1}.

\section{The clustering signature in the modified 3D numerical model}\label{appendix:r_cluster}

\begin{figure}[htp!]
\centering
\includegraphics[width=\linewidth]{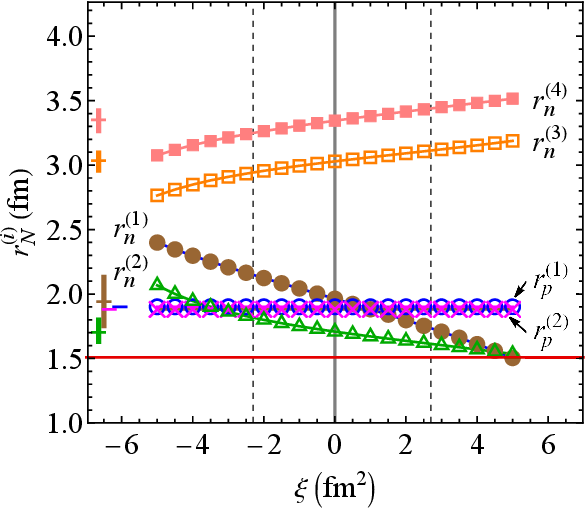}
\caption{
The CM-to-nucleon distance $r_{N}^{(i)}$ ($i$ runs through nucleons within the same nucleon species $N=n$ or $p$) in the modified 3D numerical model for the dominant two-body spin configuration of ground state $^6$He. For comparison, the red horizontal band across the plot shows $r_{N}^{(i)}$ for $^4$He nucleons, with the range of values represented by the bandwidth.}
\label{fig:he6_numer_r}
\end{figure}

To explore the signature of clustering, Fig.~\ref{fig:he6_numer_r} shows the Euclidean distance from the center of mass of $^6$He (denoted CM) to the location for each nucleon. Here, we denote the CM-to-nucleon distance $r_{p}^{(1)}$, $r_{p}^{(2)}$ for the two protons, and $r_{n}^{(1)}$ to $r_{n}^{(4)}$ for the four neutrons. We notice that, in the reasonable range of $\xi$, $r_{p}^{(1)}$, $r_{p}^{(2)}$, $r_{n}^{(1)}$ and $r_{n}^{(2)}$ exhibit overlap; in contrast, $r_{n}^{(3)}$ and $r_{n}^{(4)}$ are higher and separated from the former group by a gap of roughly 1~fm. In a classical picture, this suggests that the $\alpha$ core nucleons tend to cluster in the same orbit, whereas the two valence neutrons occupy higher orbitals.

We observe that $r_{N}^{(i)}$ display different values even for nucleons at the same shell level. For example, $r_{n}^{(1)}$ deviates from $r_{n}^{(2)}$. This is due to the fact that the left hand side of Eq.~(\ref{rij}) is fixed by the separation between nucleons, not the CM-to-nucleon distance. From Ref.~\cite{paper1}, the $\alpha$-core $np$ pairs form three triplet states and one singlet state in the dominant two-body spin configuration of the ground state. In our semiclassical picture, this means one neutron forms a singlet state with one proton, and forms a triplet state with the other proton, while the other neutron forms triplet states with both protons. Since the rms separation for a singlet state is usually different from that for a triplet state, asymmetry is induced, and thus $r_{N}^{(i)}$ values are different even for identical nucleon species belonging to the same shell.

In Fig~\ref{fig:he6_numer_r}, the left horizontal bar for each nucleon indicates the average $r_{N}^{(i)}$ over the reasonable $\xi$ range, associated by a min-max uncertainty bar for this range. We get $r_p^{(1)}=1.90$~fm and $r_p^{(2)}=1.88$~fm, with vanishingly small uncertainty. $r_{n}^{(i)}$ are  $1.94_{-0.21}^{+0.21}$~fm,  $1.70_{-0.09}^{+0.11}$~fm,  $3.03_{-0.09}^{+0.08}$~fm and  $3.35_{-0.10}^{+0.09}$~fm for $i=1$ to 4, respectively. Hence, the average CM-to-nucleon distances for protons and neutrons are 1.89~fm and 2.60(11)~fm, respectively, which are consistent with $r_p=1.89(1)$ and $r_n=2.61(3)$ calculated in Ref.~\cite{paper1}. For $^4$He, the values of $r_{N}^{(i)}$ are in the range of 1.5051~fm to 1.5114~fm, in agreement with $r_p\approx r_n\approx 1.51$~fm \cite{paper1}. Therefore, the semiclassical 3D model reproduces $r_p$ and $r_n$ that serve as indications of halo structure in the original \textit{ab initio} calculations.

\section{Illustration of the modified 3D numerical model}\label{appendix:3d_model}
\begin{figure}[htp]
  \centering
\includegraphics[width=0.4\textwidth]{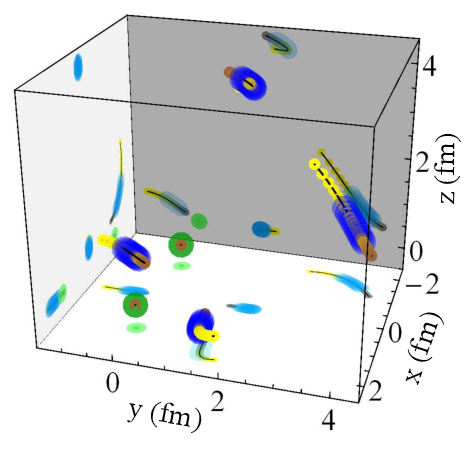}
  \caption{The modified 3D numerical model for the ground state dominant two-body spin configuration of $^6$He. For each nucleon, the connected points show a gradual transition from yellow ($\xi=-5\text{~fm}^2$) to brown ($\xi=5\text{~fm}^2$). The green (blue) corona is to signify the positions of protons (neutrons) with $-2.3\text{~fm}^2\leq\xi\leq2.7\text{~fm}^2$. The shadows on perpendicular walls indicate the projections for the nucleon positions.}
  \label{fig:model_modified}
\end{figure}

The resulting positions of the nucleons for the modified 3D numerical model (Sec.~\ref{sec_model3}) are illustrated in Fig.~\ref{fig:model_modified}. Throughout the reasonable range of $\xi$, the four $\alpha$ core nucleons nearly stay in the same plane. We also notice that as the separation of the two valence neutrons increases, the separation of the $\alpha$ core neutrons decreases. 

\newpage

\end{document}